\documentclass[aps,prl,twocolumn]{revtex4}
\usepackage{graphics}

\begin{document}

\title{Self-compensation in manganese-doped ferromagnetic semiconductors}

\author{Steven C. Erwin}
\author{A. G. Petukhov}
\altaffiliation[Permanent address: ]{Dept.\ of Physics,
South Dakota School of Mines and Technology, Rapid City, SD 57701}
\affiliation{Center for Computational Materials Science, 
Naval Research Laboratory, Washington, D.C. 20375}

\date{\today}

\begin{abstract}
We present a theory of interstitial Mn in Mn-doped ferromagnetic
semiconductors.  Using density-functional theory, we show that under
the non-equilibrium conditions of growth, interstitial Mn is easily
formed near the surface by a simple low-energy adsorption pathway.  In
GaAs, isolated interstitial Mn is an electron donor, each compensating
two substitutional Mn acceptors. Within an impurity-band model, 
partial compensation {\em promotes} ferromagnetic order below
the metal-insulator transition, with the highest Curie temperature
occurring for 0.5 holes per substitutional Mn.
\end{abstract}


\maketitle

Ferromagnetism in dilute magnetic semiconductors is generally believed
to be mediated by carriers---electrons or holes---originating from the
magnetic dopants themselves.  For example, an isolated Mn impurity in
GaAs can substitute for Ga and contribute one hole, which is weakly
bound to its acceptor core \cite{schneider87a}. GaAs samples with
Mn dopant concentrations in the range 1--2\% are ferromagnetic
insulators, while samples in the range 3--6\% are ferromagnetic metals
\cite{ohno99a}. In the metallic phase the nominal hole concentration, 
$p$, is in principle equal to the number of Mn atoms per unit
volume. Measured hole concentrations are much smaller, by factors
ranging from $\sim$3 for MnGaAs \cite{ohno99a} to $\sim$10 or more for
MnGe \cite{park01a}.  In most theories of ferromagnetism in dilute
magnetic semiconductors, reduced hole concentrations suppress the
Curie temperature
\cite{dietl00a,konig01a,sanvito02a}.  The reverse scenario---raising
the Curie temperature by increasing the hole concentration---is
therefore of great current interest.  For current theoretical reviews
see Refs.~\cite{konig01a,sanvito02a}.

Recent experiments show a strong correlation between Curie
temperature, carrier concentration, and the fraction of Mn found at
interstitial sites \cite{yu02a}.  In this paper we address several
questions not yet settled by experiment. (1) By what mechanism are
interstitials formed, given that their calculated formation energies
are considerably higher than substitutionals? (2) What determines the
relative abundance of interstitials and substitutionals? (3) Under
what (doping) conditions do interstitials act as compensators? (4)
What role does compensation play in the ferromagnetism?  

To answer these questions, we use density-functional theory (DFT) to
establish the following: (i) During the MnGaAs growth, Mn adatoms
follow a very simple low-energy pathway to directly form interstitial
Mn near the surface.  (ii) The deposition of additional As converts
some of these interstitials to substitutional sites.  (iii) The
remaining interstitial Mn atoms act as donors, each compensating two
substitutional acceptors. Finally, we show that within an
impurity-band model, ferromagnetism below the metal-insulator
transition is most favorable---in the sense of the highest Curie
temperature---for 0.5 holes per substitutional Mn.

For MnGaAs grown by MBE, recent channeling
Rutherford backscattering experiments show that as much as $\sim$15\%
of the total Mn may be interstitial \cite{yu02a}.  An
open theoretical question is how interstitial Mn might be formed,
under what conditions, and in what concentration.  For a system in
thermodynamic equilibrium, the concentration of each impurity species
is determined by its formation energy. Our calculations (described
below) show that the equilibrium concentration of interstitial Mn is
negligible.  But thermodynamic equilibrium can only be achieved if
local metastable configurations can be readily overcome. We find that
the energy barriers separating interstitial and substitutional
configurations do not satisfy this criterion. Hence, equilibrium
thermodynamics is not a reliable guide for studying the formation of
interstitial Mn.

To investigate the incorporation of Mn under non-equilibrium
conditions, one must identify specific reaction pathways and calculate
their energy barriers. At low Mn concentrations, the growth of MnGaAs
at low temperatures is governed by the potential-energy surface for
individual Mn adatoms adsorbing and diffusing on the GaAs surface. We
confine our attention to the GaAs(001) surface, the standard
orientation for growing MnGaAs. Under As-rich conditions, the
GaAs(001) surface can have several different reconstructions, all of which
contain surface As dimers as a common building block. Thus we studied
adsorption on a chemically reasonable model surface
consisting of five layers of bulk GaAs plus a dimerized As top layer;
the bottom layer was passivated. We calculated total energies and
forces using pseudopotentials and the generalized-gradient approximation
\cite{kresse93a,kresse96a}.  To identify the lowest energy adsorption
site we placed the Mn adatom at various surface sites in a (2$\times$2)
supercell, then relaxed the height of the adatom and the positions of
all atoms in the top five layers of the surface.  By far the most
stable adsorption site was the As-dimer bridge site, i.e., equidistant
from the two As atoms in a dimer.

%% FIG 1
\begin{figure}
\resizebox{8.5cm}{!}{\includegraphics{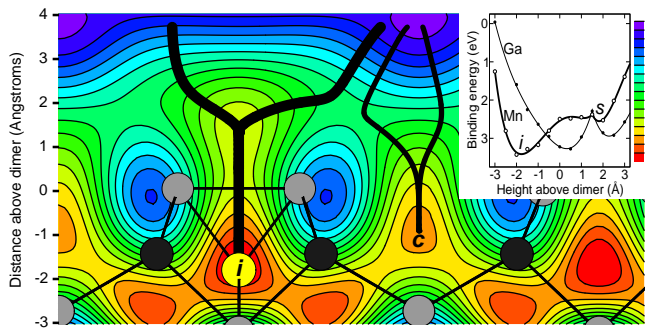}}
\caption{(color) Potential-energy surface for Mn adsorption on GaAs(001),
plotted in a plane normal to the surface and containing the As surface
dimer.  The minimum energy adsorption site is the subsurface
interstitial site labeled $\textsl{i}$; the corresponding surface
geometry is shown (light gray for As, dark gray for Ga, yellow for
Mn).  Typical adsorption pathways funnel Mn adatoms to this
interstitial site (heavy curves) or to a cave site, $\textsl{c}$
(light curves). Inset: Binding energy of a Mn adatom centered
on the As-dimer; for comparison, results are also
shown for a Ga adatom. When additional As is deposited, the metastable
Mn site, $\textsl{s}$, becomes more favorable and leads to
partial incorporation of substitutional Mn.}
\end{figure}

To identify specific adsorption pathways, we calculated the total
energy on a grid of Mn positions in a plane normal to the surface and
containing the As dimer.  The resulting potential-energy surface and
several possible adsorption pathways are shown in Fig.~1.  Beginning
from a height of several \AA, Mn adatoms are attracted to the surface
and at low kinetic energies will be ``funneled'' into one of two
adsorption channels.  The primary channel leads to the As-dimer bridge
site.  As shown in the inset to Fig.~1, there are two locally stable
adsorption positions above the As dimer: a shallow metastable site 2
\AA\ above the surface and a stable site 2 \AA\ {\em below} the
surface and more favorable by 0.8 eV. These two minima are separated
by a small barrier of 0.2 eV, associated with the opening of the As
dimer, which is easily overcome at normal growth temperatures.  (In
comparison, Ga adatoms also have two locally stable adsorption sites,
in agreement with Ref.~\cite{kley97a}. However, for Ga the
interstitial position is very unfavorable, and the stable site near
the surface layer ultimately leads to completely substitutional
incorporation \cite{kratzer02a}.)  Isolated Mn adatoms that are
steered into this primary channel will generally reach their
equilibrium position below the As dimer.  This position corresponds to
a bulk interstitial site with four neighboring As atoms.  With the Mn
atom at this position, the As dimer above it remains intact with a
slightly strained bond. A secondary adsorption channel leads, with no
energy barrier, to the cave site between adjacent dimer rows; upon
continued growth of GaAs this site will also correspond to a bulk
interstitial site. Thus we have identified a set of very low energy
pathways which initially steer isolated Mn adatoms to interstitial sites.

Although the energetics of Mn on clean GaAs leads to incorporation
only at interstitial sites, the subsequent deposition of additional
GaAs can change this situation and lead to partial substitutional
incorporation.  To illustrate this, we computed the change in Mn
binding energy resulting from an additional half monolayer of Ga or As
on the surface.  In the presence of either a Ga or As adlayer, the
surface As-dimer bond breaks and the metastable Mn site above it then
corresponds to a substitutional site in the zincblende lattice. The
adlayer completes the fourfold coordination of this site and hence may
change its stability relative to the interstitial site.  The change
depends on the type of adlayer: for a Ga adlayer, the Mn interstitial
site remains 0.8 eV more favorable than the substitutional site, but
for an As adlayer this difference is reduced to just 0.1 eV.  This
reduction reflects the crossover from the case of Mn adatoms on clean
GaAs (which favors interstitials) to the thermodynamic limit of
isolated Mn in bulk GaAs (which favors substitutionals), and suggests
that during the growth Mn will be incorporated at both interstitial
and substitutional sites.

The electrical activity of a Mn impurity is determined by its
formation energy as a function of its charge state.  We used DFT to
calculate the formation energies of substitutional and interstitial Mn
using supercells containing 54 atoms, with convergence checks using
128 atoms. We fixed the chemical potentials to correspond to the
As-rich, Mn-rich conditions normally used in growth.  Fig.~2 shows the
resulting formation energies, as a function of the Fermi level,
for charge states that are stable within the
GaAs bandgap. Substitutional Mn is an acceptor, with stable 1$-$ and
neutral charge states. The theoretical acceptor ionization energy is
100 meV, in good agreement with the experimental value of 113 meV
\cite{schneider87a}. Interstitial Mn is a deep donor, with stable
2$+$, 1$+$, and neutral charge states; its donor levels have not been
measured experimentally.  For $p$-type material, corresponding to the
Fermi level near the valence band edge, each interstitial will
compensate two substitutionals, in agreement with previous 
calculations
\cite{chang01a,masek02a}.  
The resulting hole concentration is $p=(x-2y)(4/a^3)$, where $x$ and
$y$ are the number of substitutionals and interstitials per Ga site,
and $a$ is the the GaAs lattice constant.

%% FIG 2
\begin{figure}
\resizebox{7cm}{!}{\includegraphics{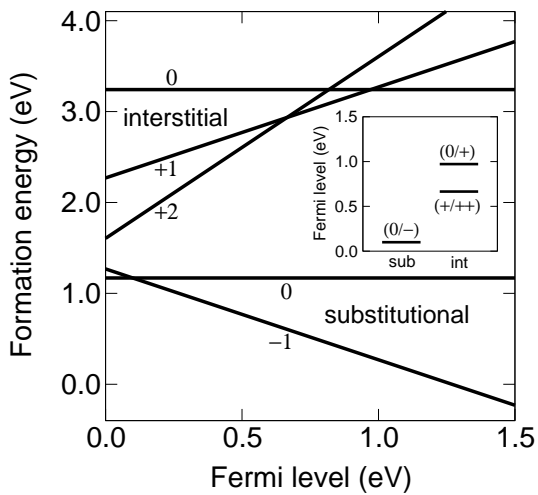}}
\caption{Formation energies of isolated Mn impurities in GaAs, for different
charge states as a function of the Fermi level. Inset: Transition
levels (crossing points of the formation energies)
between stable charge states of substitutional and interstitial Mn
impurities.}
\end{figure}

The consequences of partial compensation for ferromagnetism are
different above and below the metal-insulator transition. In the
metallic phase the Mn-induced impurity states broaden and merge with
the GaAs valence band.  Most current theories therefore consider the
carriers as free holes in the valence band. In this picture the Curie
temperature is proportional to $p^{1/3}$, and hence compensation can only
reduce the Curie temperature.  But the role of compensation is less
obvious for Mn concentrations between 1 and 2\%, where ferromagnetism
persists experimentally even though transport data clearly show the
samples to be insulating \cite{ohno99a}.  The nature of this
insulating state is not currently understood. If the Coulomb $U$ is
larger than the impurity band width, then this state is a Mott-Hubbard
insulator and an impurity-band model with localized carriers is an
appropriate starting point \cite{berciu01a}.  Below we show that
within such a model, the compensation affects the magnetic
interactions in a surprising way: the maximum Curie temperature is
obtained not for zero compensation, but rather for a partially
compensated system with 0.5 holes per substitutional Mn.

We derive this result starting from a Zener
double-exchange Hamiltonian for the motion of holes in a narrow
impurity band \cite{izyumov01a},
\begin{equation}
\label{Hamiltonian}
H_{\text{DE}}=\sum_{i\sigma}\epsilon_i a_{i\sigma}^\dagger
a_{i\sigma}+\sum_{ij\sigma}t_{ij}
a_{i\sigma}^\dagger 
a_{j\sigma}-J_{H}\sum_i \vec s_i\cdot \vec S_i.
\end{equation}
The first term describes fluctuations in the on-site impurity levels
caused by the random distribution of charged ($A^-$) and neutral
($A^0$) Mn acceptors;  the second term describes the hopping of
holes between Mn spins; and the last term describes the exchange coupling
between holes and Mn spins.  The energies $\epsilon_i$ are distributed
within the disorder-broadened band width, $W$; $a_{i\sigma}^\dagger$ and $a_{i\sigma}$ are creation
and annihilation operators for holes with spin $\sigma$ at site $i$;
$t_{ij}$ are hopping integrals; $J_H$ is the exchange coupling
constant; $\vec s_i$ and $\vec S_i$ are the operators for the hole and
Mn ($S=5/2$) spins, respectively.

It is well established that the coupling between the hole and the Mn
spin is antiferromagnetic, $J_H < 0$, and that $S|J_H|$ is in the
range 100--250 meV \cite{dietl00a,matsukura98a}.  This is comparable
to, or even larger than, the acceptor ionization energy.  Hence the
hole at site $i$ is always antiparallel to the Mn spin, and the
Hamiltonian can be reduced to a spinless form with the hopping
renormalized to $t_{ij}\cos(\theta_{ij}/2)$, where $\theta_{ij}$ is
the angle between $\vec S_i$ and $\vec S_j$.

To second order in the hopping, $H_{\text {DE}}$
can be mapped to an effective Heisenberg
Hamiltonian describing the interaction between Mn spins,
\begin{equation}
\label{Heisenberg}
H_{\text{eff}}=\frac{1}{2}\sum_{ij}\frac{\left|t_{ij}\right|^2}{\epsilon_i-\epsilon_j}
\left(n_i-n_j\right)\left(1+
\frac{\vec S_i\cdot\vec S_j}{S(S+1)}\right),
\end{equation}
where $n_i$ is the number of holes at site $i$ (thus, $n_i$=0 and 1
correspond to $A^-$ and $A^0$ acceptors, respectively).  Two features
of $H_{\text{eff}}$ are noteworthy. First, the effective coupling
between spins falls off exponentially with their separation, $r_{ij}$,
because the hopping integrals have the form $t_{ij} =
t_0(r_{ij})\exp(-r_{ij}/a_0)$, where $t_0(r)\propto r/a_0$
for hydrogenic wavefunctions and $a_0$ is the effective Bohr radius of
the Mn acceptor \cite{shklovskii84a}. Second, because of the factor $(n_i-n_j)$,
interactions occur only between Mn acceptors in different charge
states, that is, between $A^0$ and $A^-$.  
\textit{This implies that ferromagnetism is absent in the limiting cases
of no compensation and complete compensation.}

This result can be easily understood within a simple toy model.
Consider the Hamiltonian $H_{\text{DE}}$ applied to two Mn acceptors
that share either zero, one, or two holes. In the case of zero holes
(both $A^-$ acceptors), there is no interaction at all. For one hole
($A^0$ and $A^-$), the interaction is ferromagnetic, because the hole
can lower its kinetic energy via hopping only for aligned spins.  For
two holes (both $A^0$), the interaction is always antiferromagnetic
and scales as $t^2_{ij}/J_H$.  In the real system, aligning the spins
in the absence of compensation drives the Fermi level into the gap between
spin-up and spin-down states, making ferromagnetism unfavorable.

To find the Curie temperature of the ferromagnetic phase transition,
we first note that mean-field theory is not well justified because of
the exponential dependence of the integrals $t_{ij}$ and because of
the Mn positional disorder
\cite{korenblit78a,litvinov01a}. Instead we represent the system
as a percolation network of randomly distributed sites, each occupied
by either $A^0$ or $A^-$, with interactions only between sites with
different charges. In the standard solution to this two-color
percolation problem
\cite{ioselevich95a}, the critical percolation
radius, $R_c$, depends on the solution of the one-color problem
and on the maximum eigenvalue of the connectivity matrix for the
two-color problem. In our case this leads to
\begin{equation}
\label{xic}
R_c=a \left(\frac{B_c}{(16\pi/3)x\sqrt{q(1-q)}}\right)^{1/3},
\end{equation}
where $B_c=2.7$ is the average number of bonds within the critical
radius for the one-color problem
\cite{ioselevich95a}, and $q$ is the number of holes
per substitutional Mn.  In general, $q$ must be found from the
electroneutrality condition and the energy levels in Fig.~2, but at
reasonable temperatures the approximation $q=1-2y/x$ is
extremely accurate for $y<x/2$.  Note that due to the factor
$\sqrt{q(1-q)}$, the critical radius diverges for both $y=0$ and
$y=x/2$, consistent with our earlier conclusion that ferromagnetism
vanishes at these limits.

The Curie temperature is given within percolation theory by
\begin{equation}
\label{Tcgen}
k_B T_c=\frac{1}{6}B_c\left\langle \frac{n_i-n_j}
{\epsilon_i-\epsilon_j}\right\rangle_{R_c} t_0(R_c)^2\exp{(-2R_c/a_0)},
\end{equation}
where $k_B$ is the Boltzmann constant.  The angle brackets denote the
most probable value among atoms within the critical radius.  We
estimate this value using a hydrogenic impurity wavefunction with
effective Bohr radius $a_0=7.8$ \AA~\cite{berciu01a},
and take the energy separation near the Fermi level to be the Coulomb
gap \cite{shklovskii84a,efros92a}, given by
$|\epsilon_i-\epsilon_j|\simeq e^2/\kappa r_{ij}$, where $e$ is the
electron charge and $\kappa=10.66$ is the dielectric constant of GaAs
\cite{berciu01a}. This gives the Curie
temperature simply as
\begin{equation}
\label{Tc}
k_B T_c=\frac{B_ce^2}{108\;\kappa a_0} \left(\frac{2R_c}{a_0}\right)^3 \exp(-2R_c/a_0).
\end{equation}       
We show in Fig.~3(a) the dependence of the Curie temperature on the
number of holes per substitutional Mn and, equivalently for the case
of pure self-compensation, on the fraction of interstitial Mn.  The
maximum Curie temperature is reached for $q=0.5$ holes per
substitutional Mn, corresponding to an interstitial fraction $y=x/4$.

%% FIG 3
\begin{figure}
\resizebox{5.5cm}{!}{\includegraphics{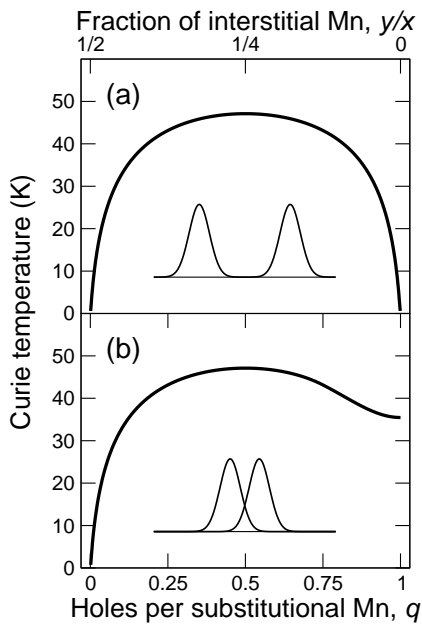}}
\caption{Dependence of Curie temperature on hole concentration for
1.5\% Mn, in two different insulating regimes.  (a) Mott-Hubbard insulator,
$U>W$. (b) Anderson localization, with the arbitrary choice
$U=0.65W$. Upper and lower Hubbard bands are shown.}
\end{figure}

The above theory ignores the orbital degeneracy of the impurity
band. This simplification is justified if the Coulomb energy, $U$, of
two holes in different orbital states is larger than both the hopping
integral and the impurity band width, $W$.  For the case when $U$ is
smaller than $W$, the resulting overlap of the two Hubbard bands will
lead to a finite density of states at the Fermi level, even without
compensation.  For Mn concentrations between 1 and 2\% the
materials are nonetheless insulating, suggesting that the states are
localized by an Anderson transition induced by strong disorder.  In
this case, hopping---and thereby also ferromagnetic coupling---can
take place not only between $A^0$ and $A^-$ but also between $A^0$ and
$A^+$ acceptors. The percolation radius is still given by
Eq.~\ref{xic}, but $q$ now must be replaced by the fraction of
neutral acceptors, which must be calculated from the density
of states. For a simple triangular density of states, the resulting
Curie temperature is shown in Fig.~3(b).
One significant new feature arises: ferromagnetism now persists even
without compensation.  Otherwise, the main features of the
Mott-Hubbard case are obtained here as well, and the maximum Curie
temperature is again reached for $q=0.5$.

\begin{acknowledgments}
This work was supported by the Office of Naval
Research, the DARPA SpinS program, and NSF Grant DMR-0071823.  We
thank C.S. Hellberg and I.I. Mazin for helpful discussions.
\end{acknowledgments}


\begin{thebibliography}{10}

\bibitem{schneider87a}
J. Schneider, U. Kaufmann, W. Wilkening, M. Baeumler, and F. K{\"o}hl, Phys.\
  Rev.\ Lett.\ {\bf 59},  240  (1987).

\bibitem{ohno99a}
H. Ohno, J. Magn.\ Magn.\ Mater.\ {\bf 200},  110  (1999).

\bibitem{park01a}
Y.~D. Park, A. Hanbicki, S.~C. Erwin, C.~S. Hellberg, J.~M. Sullivan, J.~E.
  Mattson, T.~F. Ambrose, A. Wilson, G. Spanos, and B.~T. Jonker, Science {\bf
  295},  651  (2001).

\bibitem{dietl00a}
T. Dietl, H. Ohno, F. Matsukura, J. Cibert, and D. Ferrand, Science {\bf 287},
  1019  (2000).

\bibitem{sanvito02a}
S. Sanvito, G. Theurich, and N.~A. Hill, J. Superconductivity {\bf 15},  85
  (2002).

\bibitem{konig01a}
J. K{\"o}nig, J. Schliemann, T. Jungwirth, and A.~H. MacDonald,
  cond-mat/0111314  (2001).

\bibitem{yu02a}
K. Yu, W. Walukiewicz, T. Wojtowicz, and J. Furdyna, Phys.\ Rev.\ B {\bf 65},
  201303(R)  (2002).

\bibitem{kresse93a}
G. Kresse and J. Hafner, Phys.\ Rev.\ B {\bf 47},  558  (1993).

\bibitem{kresse96a}
G. Kresse and J. F{\"u}rthmuller, Phys.\ Rev.\ B {\bf 54},  11169  (1996).

\bibitem{kley97a}
A. Kley, P. Ruggerone, and M. Scheffler, Phys.\ Rev.\ Lett.\ {\bf 79},  5278
  (1997).

\bibitem{kratzer02a}
P. Kratzer and M. Scheffler, Phys.\ Rev.\ Lett.\ {\bf 88},  036102  (2002).

\bibitem{chang01a}
Y.~H. Chang and C.~H. Park, J.\ Korean\ Phys.\ Soc.\ {\bf 39},  324  (2001).

\bibitem{masek02a}
J. Masek and F. Maca, Phys.\ Rev.\ B {\bf 65},  235209  (2002).

\bibitem{berciu01a}
M. Berciu and R.~N. Bhatt, Phys.\ Rev.\ Lett.\ {\bf 87},  107203  (2001).

\bibitem{izyumov01a}
Y.~A. Izyumov and Y.~W. Skryabin, Phys.\ Usp. {\bf 44},  109  (2001).

\bibitem{matsukura98a}
F. Matsukura, H. Ohno, A. Shen, and Y. Sugawara, Phys.\ Rev.\ B {\bf 57},
  R2037  (1998).

\bibitem{shklovskii84a}
B.~I. Shklovskii and A.~L. Efros, {\em Electronic properties of doped
  semiconductors} (Springer-Verlag, Berlin, 1984).

\bibitem{korenblit78a}
I.~Y. Korenblit and E.~F. Shender, Sov.\ Phys.\ Usp.\ {\bf 126},  233  (1978).

\bibitem{litvinov01a}
V.~I. Litvinov and N.~K. Dugaev, Phys.\ Rev.\ Lett.\ {\bf 86},  5593  (2001).

\bibitem{ioselevich95a}
A.~S. Ioselevich, Phys.\ Rev.\ Lett.\ {\bf 74},  1411  (1995).

\bibitem{efros92a}
A.~L. Efros, Phys.\ Rev.\ Lett.\ {\bf 68},  2208  (1992).

\end{thebibliography}
\end{document}